\documentclass[%
 reprint,
 amsmath,amssymb,
 aps,
]{revtex4-1}

\usepackage{graphicx}
\usepackage{dcolumn}
\usepackage{bm}

\begin{document}

\preprint{APS/123-QED}

\title{Reconstruction of $\alpha$-attractor supergravity models of inflation}

\author{Alessandro Di Marco}
\email{alessandro.di.marco@roma2.infn.it}
\author{Paolo Cabella}
\author{Nicola Vittorio}
\affiliation{%
University of Rome - Tor Vergata, Via della Ricerca Scientifica 1\\
INFN Sezione di Roma “Tor Vergata”, via della Ricerca Scientifica 1, 00133 Roma, Italy}%


\begin{abstract}
In this paper, we apply reconstruction techniques to recover the potential parameters for a particular class of single-field models, the $\alpha$-attractor (supergravity) models of inflation. 
This also allows to derive the inflaton vacuum expectation value at horizon crossing. 
We show how to use this value as one of the input variables to constrain the postaccelerated inflationary phase.
We assume that the tensor-to-scalar ratio $r$ is of the order of $10^{-3}$, a level reachable by the expected sensitivity of the next-generation CMB experiments.

\end{abstract}

\pacs{Valid PACS appear here}
\maketitle


\section{Introduction}

Different cosmological observations \cite{1,2,3,4,5} have converged by now to
support $\Lambda$CDM as the concordance model of modern cosmology.
This model, as all its possible other extensions, assumes as a paradigm the inflationary scenario.
This is needed for two reasons:
on one hand, to justify the observed flatness and isotropy of the Universe, as well as the absence of magnetic monopoles \cite{6,7,8,9,10};
on the other hand, to exploit quantum mechanisms for explaining the origin of matter \cite{11,12,13,14,15} and the production of those fluctuations responsible for the formation of the large scale structure of the Universe \cite{16,17,18,19,20,21,22,23,24,25,26,27,28}.
However, two general questions still need need to be investigated: the shape of the inflationary potential and its energy scale.
The quantities related to these features are the scalar spectral-index, $n_s$, and the tensor-to-scalar ratio amplitude, $r$.
The current estimations for these parameters provide $n_s=0.968 \pm 0.006$ and $r_{0.002}<0.07$ at the $95\%$ 
confidence level (see \cite{4,5}).
Because of this, we have now a clear idea on the health state of some specific inflationary models:
most of the minimally coupled power-law potentials are ruled out, while exponential potentials with a very flat region, seem to be favored by current data as outlined especially in \cite{4}. 
Furthermore, there is still room to look into other aspects of inflation:
the fundamental mechanism that induces the inflationary phase; the initial condition for the inflaton field, its nature and its mass $m_{\phi}$ 
\cite{29,30,31,32,33,34,35}; and the possibility for a multifield inflation and the induced non-Gaussianity in the cosmic macrowave background (CMB) fluctuations \cite{36}.
In this paper we will consider a tensor-to-scalar ratio $r \approx 10^{-3}$, 
consistent with the expected sensitivity of the next-generation CMB experiments \cite{37,38,39,40,41}.
We will show the statistical information that can be derived on a very important class of inflationary potential, the so called $\alpha$-attractor models of inflation.
This class of models can be generated in different ways, although the most advanced version emerges from the supergravity context (see refs \cite{42,43,44,45,46,47} for properties and details).
It is important to stress that $\alpha$-attractors include the first plateau-type potential, the Goncharov-Linde model \cite{48}, the Starobinsky modified gravity $R^2$ scenario \cite{49,50} and Higgs Inflation \cite{51}.
In particular we consider the E-model version of the $\alpha$-attractor class.
To reach this goal we have to reconstruct the inflationary potential.
Among the different algorithms proposed in the literature \cite{52,53,54,55,56,57,58,59,60,61,62,63}, we will focus on
the simple approach based on constraining the local shape of the potential during the pure accelerated phase.
This is done by implementing a Taylor expansion around the vacuum expectation value of the inflaton field at horizon crossing, $\phi_{*}$, and by connecting the coefficients of the expansion to the observables $n_s$ and $r$ (details in \cite{52,53}).
In its simplicity, this procedure provides a model independent estimation of the inflationary potential around $\phi_*$.
Hereafter, we show that for $\alpha$-attractor models it is possible to derive constraints also on the vacuum expectation value $\phi_*$.
Now, on one hand it is true that $\phi_*$ by itself is not important: the fundamental quantity that parameterizes the inflationary evolution and the cosmological variables is the number of e-foldings $N_*$.
On the other hand, there could be a couple of reasons to constrain $\phi_*$.
First, this value could be of interest from the particle physics point of view.
Secondly, as we shall see later, it is simpler to directly use $\phi_*$ as one of the input variables to constrain $N_*$ and so, the reheating phase.
In the following, we use the natural units of particle and cosmology $c=\hbar=k_B=1$, unless otherwise indicated.
The paper is organized as follows.
In Sec. II , we review the general properties of inflation and of the slow roll dynamics.
In Sec. III, we focus on the magnitude of $\phi_*$ in different inflationary models.
In Sec. IV, we discuss the basics of the local potential reconstruction and we apply such a procedure to evaluate the parameters of the chosen inflationary models.
Finally, the last section is dedicated to the discussion of our findings and to possible extensions of this work.

\section{Inflationary slow roll dynamics and its observables}

Inflation is defined as an early accelerated expansion phase.
Therefore, the evolution of the scale factor is almost nearly exponential $a(t)\sim e^{N}$, where $N$ is the so-called number of e-foldings.
Such a condition implies a nearly constant Hubble rate, $H(t)$, i.e., a nearly constant Hubble radius, $R_H=c/H$.
The simplest scenario for inflation involves a neutral and homogeneous scalar field $\phi$, called inflaton, that is minimally coupled to gravity with a canonical kinetic term.
When such a field dominates, inflation occurs, giving rise 
to an accelerated expansion that occurred between $10^{-35} s$ and $10^{-32} s $ after the initial singularity,
on an energy scale below the GUT scale ($E < 10^{16}$ GeV).
The inflaton field evolves accordingly to a potential $V(\phi)$, characterized by an almost flat region.
The cosmological action for early times is the following:
\begin{eqnarray}
S=\int d^4x \sqrt{-g} \left\{\frac{1}{2}M^2_p R - \frac{1}{2}g_{\mu\nu}\partial^{\mu}\phi \partial^{\nu}\phi -V(\phi)   \right\}
\end{eqnarray}
where $R$ is the Ricci scalar, $g_{\mu\nu}$ is the metric tensor, $g$ its determinant and $M_p$ the reduced Planck mass.
The inflationary equations for a FRW flat Universe in the Hamilton-Jacobi formalism, are
\begin{eqnarray}\label{eqn: equazione HJ}
V(\phi)&=&3M^2_p H^2(\phi) - 2 M^2_p  H'(\phi)
\end{eqnarray}
\begin{eqnarray}\label{eqn: fieldeq}
\dot{\phi}&=&-2 M^2_p H'(\phi)                
\end{eqnarray}
where $'$ denotes derivative with respect to the scalar field. It is required that the sign of $\dot{\phi}$ does not change, in order to have a monotonic evolution of the field.
Therefore, without loss of generality, we can choose $\dot{\phi}<0$ so that $H'\ge 0$, or the opposite case.
Inflation starts when the inflaton moves slowly through the almost flat region of the potential.
In this phase, the kinetic term in the action is negligible with respect
to the potential:
\begin{equation}\label{eqn: piatto}
\partial_{\mu}\phi\partial^{\mu}\phi \ll V(\phi)
\end{equation}
Afterwards, when the inflaton reaches the potential global minimum, the reheating phase can start (see \cite{11,12,13,14,15} for more details).
In this phase, the field oscillates and decays producing entropy.\\
Once the functional form of the potential is given, one can describe the inflaton dynamics via the Potential Slow Roll Parameters (PSRP), defined as follows:
\begin{eqnarray}\label{eqn: psrp }
\epsilon_V(\phi)=\frac{M^2_p}{2} \left(\frac{V'(\phi)}{V(\phi)} \right)^2,  \quad \eta_V(\phi)= M^2_p \left( \frac{V''(\phi)}{V(\phi)} \right)
\end{eqnarray}
or alternatively, by the Hubble Slow Roll Parameters (HSRP)
\begin{eqnarray}\label{eqn: hsrp}
\epsilon(\phi)=2M^2_p\left( \frac{H'(\phi)}{H(\phi)}  \right)^2, \quad   \eta(\phi)=2M^2_p\left( \frac{H''(\phi)}{H(\phi)}  \right)
\end{eqnarray}
To first order,
\begin{eqnarray}\label{eqn: rel}
\epsilon \simeq \epsilon_V, \quad\quad \eta \simeq \eta_V - \epsilon_V.
\end{eqnarray}
For an exhaustive discussion on these two formalisms, on their properties and their relation we refer to reference \cite{64}.
Typically, inflation occurs when $\epsilon(\phi) < 1 $ (or $\epsilon_V(\phi) \ll 1$) and finishes when $\epsilon \sim 1$.\\
Inflation also provides a solution for the origin of primordial perturbations.
In the inflationary universe there are quantum fluctuations of thermal type with temperature equal to the Gibbons-Hawking temperature $T_H=H/2\pi$.
These thermal fluctuations allow to treat the inflaton field as a quantum field $\hat{\phi}(x,t)$
with zero mean value in a macro time scale.
Such a condition implies fluctuations on the stress-energy tensor and then, on the metric tensor.
The cosmic acceleration due to inflation stretches fluctuations up to astronomical scales.
At this stage, fluctuations freeze-out and become classical metric perturbations.
At the end of inflation, the Hubble radius starts to grow, catching those perturbations that will produce the anisotropies of
the CMB and the formation of the large scale structures.\\
Let us now consider the cosmological perturbation field $\delta(x,t)$, with power spectrum and spectral-index
\begin{equation}
P(k)=\frac{k^3}{2 \pi^2}|\delta(k)|^2 
\end{equation}
\begin{equation}
n(k)=\frac{d P(k)}{d \ln k}
\end{equation}
where $k$ is the wavenumber.
The first one describes the presence of the perturbation on a given scale $k$ while 
the second describes the variation of $\delta(k)$ with respect to the scale.
Because of the homogeneity and isotropy of the FRW background, one can decompose the perturbations in scalar, vectors and tensors modes (SVT decomposition).
In particular, inflation excites only the scalar and tensor modes.
The most familiar form of the power spectrum for the scalar and tensor sector, to first order in the slow roll parameters, are
\begin{eqnarray}\label{eqn: osservabili1}
P_s(k)= \frac{1}{8\pi^2  M_p^2}\frac{H^2}{\epsilon}\Bigl|_{k=aH}  
\end{eqnarray}
\begin{eqnarray}\label{eqn: osservabili2}
P_t(k)=\frac{2}{\pi^2}\frac{H^2}{M^2_p}\Bigl|_{k=ah}
\end{eqnarray}
The corresponding spectral indices are defined as follows:
\begin{eqnarray}
n_s=1-4\epsilon+2\eta\\
n_t=-2\epsilon
\end{eqnarray}
and the tensor-to-scalar ratio of the perturbation amplitudes is
\begin{eqnarray}\label{eqn: ratio}
r=\frac{P_s(k)}{P_t(k)}=16\epsilon=-8n_t.
\end{eqnarray}
Refs in \cite{16,17,18,19,20,21,22,23,24,25} while for the SVT-decomposition, see \cite{26,27,28}.
The quantities $n_s$ and $r$ can be computed for any given theoretical model and, then, compared with the ones estimated by the experiments.
To do this, one can use Eq.($\ref{eqn: rel}$) to express $n_s$ and $r$ in terms of the PSRP, providing the following familiar relations:
\begin{eqnarray}\label{eqn: psrpobs}
n_s\sim1-6\epsilon_V (\phi) + 2\eta_V (\phi), \quad r\sim16\epsilon_V(\phi)
\end{eqnarray}
Note, that these quantities are functions of the scalar field because of Eq.($\ref{eqn: psrp }$) or Eq.($\ref{eqn: hsrp}$).
Their magnitudes are evaluated by setting $\phi=\phi_*$, where $\phi_*$ is the value of inflaton field at horizon crossing epoch: 
\begin{equation}\label{eqn: def}
n_s=n_s(\phi_*), \qquad r=r(\phi_*)
\end{equation}
In the following Section, we discuss how to compute such a value.

\section{The classical trajectory: relation between $\phi_{*}$ and $N_*$}

\begin{figure}[htbp]
\centering
\includegraphics[width=8.6cm, height=6cm]{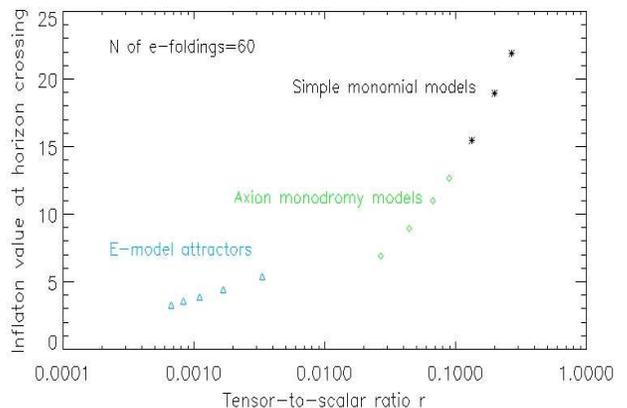}
\caption{Inflaton field values at horizon crossing vs. tensor-to-scalar ratio for: monomial models with, $n=2,3,4$ \cite{8};$\alpha$-attractor models with $\alpha=1/5,1/4,1/3,1/2,1$ \cite{42,43,44,45,46,47}; axion monodromy models with $n=2/5,2/3,1,4/3$ \cite{66,67,68}. As expected, the value of the inflaton field increases if one moves toward large field scenarios.}
\label{fig:inflaton}
\end{figure}

The inflaton dynamics is quite interesting.
In principle, once a particular potential is chosen, one can follow numerically the evolution of the function $\phi(t)$ by solving the system of Eq.($\ref{eqn: equazione HJ}$) and Eq.($\ref{eqn: fieldeq}$).
However, if the slow roll condition is satisfied, one can simplify the system with the following formula \cite{65}:
\begin{eqnarray}\label{eqn: eqofmotion}
\Delta N=\frac{1}{M_p} \int_{\Delta\phi} d\phi \frac{1}{\sqrt{\epsilon_V(\phi)}}
\end{eqnarray}
where the inflaton field can have, as discussed above, either positive or negative sign in its time derivative. Here, $\Delta\phi=|\phi-\phi_{end}|$ is the range of variation of the scalar field up to the final value $\phi_{end}$ and $\Delta N$ the related number of e-foldings.
The solution of this integral represents the classical trajectory of motion that we can rewrite as
\begin{eqnarray}\label{eqn: solution}
\phi=\phi (\phi_{end},\Delta N)
\end{eqnarray}
The inflationary trajectory is characterized by different phases which are worthy of attention: the initial conditions for both inflation and the cosmological fluctuations; the epochs of the horizon crossing and of the end of inflation.
The end of inflation is quite simple to evaluate.
In fact, it is sufficient to solve the algebraic broken-inflation condition $\epsilon(\phi_{end})=1$ to get
the possible $\phi_{end}$'s values.
While the fundamental mechanism that induces the inflationary phase is actually unknown,
one can still say something about the initial condition for inflation, $\phi_0$.
In particular, in the case of large field models, the inflationary trajectory is a ``local" attractor solution in the $\phi_0$-space, as summarized by Brandenberger in \cite{29}.
Such evidence suggests that the subsequent physical events, as the generation of cosmological perturbations, does not depend explicitly from $\phi_0$ \cite{30,31,32}.
On the other hand, the generation of cosmological fluctuations in the inflationary background occurs at an epoch commonly associated with the Bunch-Davies vacuum condition, which is an attractor in the (state) space.
This epoch it is also related to the so-called ``trans-planckian" problem, as again suggested in \cite{29}.
Finally, the horizon-crossing of cosmological fluctuations occurs when the inflaton field explores the almost plateau region of the effective potential: as seen before, when the potential term is dominant
[see Eq.($\ref{eqn: piatto}$)], the value of scalar field remains substantially the same, say $\phi_{*}$ say.
The order of magnitude of $\phi_{*}$ depends on the inflationary potential $V(\phi)$ and from the number of e-foldings $N_{*}$ before the end of inflation.
From Eq.($\ref{eqn: solution}$), we have
\begin{eqnarray}\label{eqn: dep}
\phi_*=\phi_*(\phi_{end},\beta_i,N_*)
\end{eqnarray}
where $\beta_i$ are the parameters describing the specific potential function.
As shown in Eq.($\ref{eqn: def}$), the knowledge of $\phi_*$ is useful for calculating $n_s$ and $r$. 
However, $\phi_*$ depends on $N_*$ through Eq.($\ref{eqn: dep}$).
Then both $n_s$ and $r$ can be explicitly calculated once the fundamental parameter $N_*$ is given.
The most common prescription for an order-of-magnitude evaluation of $\phi_*$ (and also $n_s$ and $r$) requires $N_*=60$.
In Fig.$\ref{fig:inflaton}$, we show a nonexhaustive plot for $\phi_*$ (in units of $M_p$) in terms of the predicted $r$, for some one-parameters inflationary models.
In particular, we present qualitative results for single power-law models, E-model version of the $\alpha$-attractor class (see Sec. I for references) and axion monodromy inflation \cite{66,67,68}.

\section{Reconstructing the $\alpha$-attractor supergravity models from next-generation CMB experiments}

\begin{figure}[htbp]
\centering
\includegraphics[width=8.6cm, height=6cm]{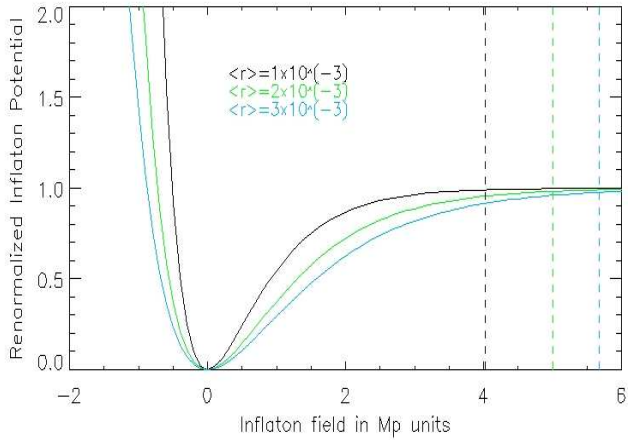}
\caption{Shape of the potentials normalized to the energy density $\Lambda^4$ and the relative inflaton value at horizon exit: the solid lines represent the three ``mean" potential curves for the computed simulations. When the mean value of $r$ increases, the curve is less steep. The dashed lines represent the three ``mean" value of $\phi_*$ at horizon crossing. When $r$ increases $\phi_*$ is always moved farther.}
\label{fig: shape}
\end{figure}

The simplest version of the potential reconstruction technique is based on a local constraint on the shape of the inflationary potential. In fact, during inflation \cite{52}:
\begin{itemize}
\item The value of the inflaton field is approximately constant,
since $\dot{\phi}^2 \ll V(\phi)$, and
\item The observable modes are stretched out over the Hubble radius, $R_H$, when $N_*\sim 60$.
\end{itemize}
Therefore, it is possible to expand the potential around $\phi_*$, the value of the inflaton field at the horizon crossing: 
$$
V(\phi)=V(\phi_*)+V'(\phi_*)(\phi-\phi_*)+\frac{1}{2}V''(\phi_*)(\phi - \phi_*)^2+... . 
$$
At this point, one can write the coefficients of this expansion in terms of the slow roll parameters and, then, with respect to the observable quantities, $n_s$ and $r$.
The weights of the polynomial form are given by the Hamilton-Jacobi equation [cf. Eq.($\ref{eqn: equazione HJ}$)].
The expansion up to the second order in $\Delta \phi$ is given by \cite{52}
\begin{eqnarray}\label{eqn: sviluppo low order}
V(\phi)=\Lambda^4\left[ 1+ d_1 \left(\frac{\Delta\phi}{M_p}\right) + \frac{1}{2}d_2 \left(\frac{\Delta\phi}{M_p}\right)^2  + ...  \right]
\end{eqnarray}
where, to first order in $n_s$ and $r$, one has
\begin{eqnarray}
\Lambda^4=\frac{3}{2}\pi^2 M^4_p P_s(k) r
\end{eqnarray}
and
\begin{eqnarray}\label{eqn: coefficenti norm}
d_1=\frac{1}{2}\sqrt{\frac{r}{2}}, \quad d_2=\frac{1}{3}\left[ 9\frac{r}{16} - \frac{3}{2}(1-n_s) \right]
\end{eqnarray}
Note that, 
the $d_i$ are dimensionless quantities, as well as the ratio $\Delta\phi/M_p$.
These definitions provide a model-independent constraint on the shape of the inflaton potential,
as they are directly connected with the first and second order derivatives of the potential.  
This formalism has been used for example by \cite{53}, for comparing theory with observations.
Even if further approaches for the reconstruction problem have been discussed in the literature (refs in Sec. I), here we want to use this local analytical approach to constrain the parameters for a class of inflationary potentials.\\
The general recipe is the following.
Let us consider a specific model of inflation with a potential $V_{\beta_i}(\phi)$, where $\beta_i$ is the set of parameters that modulates the potential function. 
We can expand this potential up to second order around $\phi_*$:
\begin{eqnarray}
V(\phi) = \Lambda^4\left[ 1+ c_1 \left(\frac{\Delta\phi}{M_p}\right) + \frac{1}{2}c_2 \left(\frac{\Delta\phi}{M_p}\right)^2  + ...  \right]
\end{eqnarray}\label{eqn: taylor}
The coefficients $c_1$ and $c_2$ are both functions of the inflaton value, $\phi_*$, and of the free parameters, $\beta_i$:
$c_1=c_1(\phi_*,\beta_i)$ and $c_2=c_2(\phi_*,\beta_i)$. 
By comparing the model-independent and the model-dependent expansion, we have
\begin{eqnarray}\label{eqn: match}
c_1=d_1, \quad\quad c_2=d_2.
\end{eqnarray}
Using these relations, we can derive predictions for $\phi_*$ and $\beta_i$ in the form: $\phi_*=\phi_*(n_s,r)$, $\beta_i=\beta_i(n_s,r)$.
The functional dependency of $d_1$ and $d_2$ on $n_s$ and $r$ is strongly model dependent.
So, there may be cases in which it is not possible to write 
both $\phi_*$ and $\beta_i$ in terms of the cosmological observables.\\
An interesting class of inflationary models is the so-called $\alpha$-attractor class (see Sec. I for details and references).
In particular, the E-model attractors are characterized by the following standard function:
\begin{eqnarray}
V(\phi)=\Lambda^4\left(1-e^{-b \phi/M_p}\right)^2, \quad b=\sqrt{\frac{2}{3\alpha}}
\end{eqnarray}
In the supergravity framework, the parameter $\alpha$ is related to the K{\"a}lher curvature of the inflaton's scalar manifold:
\begin{eqnarray}\label{eqn: kaler}
R_K=-\frac{2}{3\alpha}
\end{eqnarray}
This is a fundamental parameter in the framework of attractor models.
Let us now compute the quadratic Taylor expansion of $V$:
\begin{eqnarray}\label{eqn: scalarpot exp}
V(\phi)\simeq \Lambda^4 \left[ c_0+ c_1 \left(\frac{\Delta\phi}{M_p}\right) + \frac{1}{2}c_2 \left(\frac{\Delta\phi}{M_p}\right)^2         \right].
\end{eqnarray}
where
\begin{equation}\label{eqn: c0}
c_0 = 1-2e^{-b\phi_*/M_p} \sim 1
\end{equation}
\begin{equation}\label{eqn: c1}
c_1 = 2be^{-b\phi_*/M_p}
\end{equation}
\begin{equation}\label{eqn: c2}
c_2 = -2b^2 e^{-b\phi_*/M_p}
\end{equation}
with $\phi_*/M_p\gg 1$.
We can use these relations to evaluate $\phi_*$ and $\alpha$ from a given CMB experiment.
From Eq.($\ref{eqn: match}$) it follows that
\begin{eqnarray}\label{eqn: estb}
\frac{d_2}{d_1}=-b
\end{eqnarray}
Moreover, from Eq.($\ref{eqn: c1}$), we get
\begin{eqnarray}\label{eqn: estphi}
\frac{\phi_*}{M_p}(b,d_1)=-\frac{1}{b} \ln \left( \frac{d_1}{2b} \right).
\end{eqnarray}
These equations provide information on the inflationary models, given $n_s$ and $r$ from CMB data.
Since, current CMB experiments still do not provide a measurement on $r$, this approach is by now not very effective.
However, the situation should rapidly change in the near future, with a strong improvement on the knowledge of the tensor-to-scalar ratio.
Having this in mind, we discuss what kind of constraints we may have on $\alpha$-attractor models, assuming that next-generation of CMB experiments will be able to probe a tensor-to-scalar ratio of the order of $r\sim 10^{-3}$.\\
In the following, we simulate values of $n_s$ and $r$, randomly extracted from a gaussian multivariate distribution of the form
\begin{equation}\label{eqn: gauss}
\mathcal{G}(n_s,r)=\frac{1}{\sqrt{4\pi^2\sigma^2_{n_s}\sigma^2_{r}(1-\rho^2)}} \exp{\left(-\frac{Q^2}{2}\right)}
\end{equation}
Here
\begin{equation}
Q^2=\frac{1}{1-\rho^2}\left[\frac{(n_s-\mu_{n_s})^2}{\sigma_{n_s}^2} + \frac{(r-\mu_{r})^2}{\sigma_{r}^2} 
-2\rho\frac{(n_s-\mu_{n_s})(r-\mu_{r})}{\sigma_{n_s}\sigma_{r}}\right]
\end{equation}
where $\mu_{n_s},\mu_{r}$ and $\sigma_{n_s},\sigma_{r}$ are mean and rms values of the scalar spectral-index and the tensor-to-scalar ratio, respectively, while $\rho$ is the correlation coefficient.
In particular, we use (consistentely with current the PLANCK data) the values 
$\mu_{n_s}=0.968$ and $\sigma_{n_s}=0.006$, while for $r$ we choose three different values ($\mu_r=0.001,0.002,0.003$) with $\sigma_r=0.0001$.
The correlation coefficient between $n_s$ and $r$ is fixed to be $\rho=0.1$.
We extract from the distribution of Eq.(\ref{eqn: gauss}) pairs of values for $n_s$ and $r$.
For each extraction, we reconstruct the coefficients $d_1$ and $d_2$ from Eq.($\ref{eqn: coefficenti norm}$). 
Then, we use Eq.($\ref{eqn: estb}$) and Eq.($\ref{eqn: estphi}$) to estimate $\alpha$, $\phi_*$ and $R_K$, from a sample of $\simeq 10^4$ draws.
We note that, increasing $r$, the shape of the inflaton potential gets smoother, pushing $\phi_*$ to larger values, as shown in Fig.$\ref{fig: shape}$.


\begin{table}[htbp]
\begin{ruledtabular}
\begin{tabular}{lcdr}
\textrm{$r$}& \textrm{$d_1$ mean value}& \multicolumn{1}{c}\textrm{$d_1$ $1$-$\sigma$ value}\\
\colrule
0.001                     &     0.01117                          &     0.00056 \\                       
0.002                     &    0.01581                           &    0.00039  \\                       
0.003                     &   0.01936                            &    0.00032  \\                       
\end{tabular}
\end{ruledtabular}
\caption{Simulation results for the coefficients $d_1$ of the Taylor expansion. As we can see, the $1$-$\sigma$ value increases as the mean value of $r$ gets larger.}
\label{tab: I}
\end{table}

\begin{table}[htbp]
\begin{ruledtabular}
\begin{tabular}{lcdr}
\textrm{$r$}& \textrm{$d_2$ mean value}& \multicolumn{1}{c}\textrm{$d_2$ $1$-$\sigma$ value}\\
\colrule
0.001                     &    -0.01583                          &     0.00302 \\                       
0.002                     &    -0.01564                          &     0.00302  \\                       
0.003                     &   -0.01545                           &    0.00302 \\                       
\end{tabular}
\end{ruledtabular}
\caption{Simulation results for the coefficients $d_2$ of the Taylor expansion: the resulting $1$-$\sigma$ converges to the same value up to the 5th decimal place.
The negative sign suggests that the shape of the inflationary potential about the horizon crossing moment is locally described by a parabola which opens downward.}
\label{tab: II}
\end{table}

\begin{table}[htbp]
\begin{ruledtabular}
\begin{tabular}{lcdr}
\textrm{$r$}& \textrm{$\phi_*$ mean value}& \multicolumn{1}{c}\textrm{$\phi_*$ $1$-$\sigma$ value}\\
\colrule
0.001                     &     4.02                          &     0.70 \\                       
0.002                     &    5.02                           &    0.85  \\                       
0.003                     &   5.68                            &    0.95  \\                       
\end{tabular}
\end{ruledtabular}
\caption{Simulation results for the vacuum expectation value of the scalar field in the (supergravity) $\alpha$-attractor models. 
The table shows an increasing on the
uncertainty $\sigma$ as the mean value of $r$ increases.}
\label{tab: III}
\end{table}

\begin{table}[htbp]
\begin{ruledtabular}
\begin{tabular}{lcdr}
\textrm{$r$}& \textrm{$p$ mean value}& \multicolumn{1}{c}\textrm{$p$ $1$-$\sigma$ value}\\
\colrule
0.001                     &     -1.07                          &     0.40 \\                       
0.002                     &    -0.34                           &    0.40  \\                       
0.003                     &   0.08                            &    0.41  \\                       
\end{tabular}
\end{ruledtabular}
\caption{Simulation results for the $p=\ln\alpha$ parameter. In this case the uncertainty on the parameter is (by an large) the same up to the 2th decimal place, for all the three cases.}
\label{tab: IV}
\end{table}

\begin{table}[htbp]
\begin{ruledtabular}
\begin{tabular}{lcdr}
\textrm{$r$}& \textrm{$R_k$ mean value}& \multicolumn{1}{c}\textrm{$R_k$ $1$-$\sigma$ value}\\
\colrule
0.001                     &     -2.09                          &     0.79 \\                       
0.002                     &    -1.01                           &    0.38  \\                       
0.003                     &   -0.66                            &    0.25  \\                       
\end{tabular}
\end{ruledtabular}
\caption{Simulation results for the scalar K{\"a}lher curvature. Here, the $1$-$\sigma$ value gets larger as $r$ increases. In particular, it gets smaller as the value of $R_{K}$ become larger.}
\label{tab: V}
\end{table}

\begin{figure}[htbp]
\centering
\includegraphics[width=8.6cm, height=7cm]{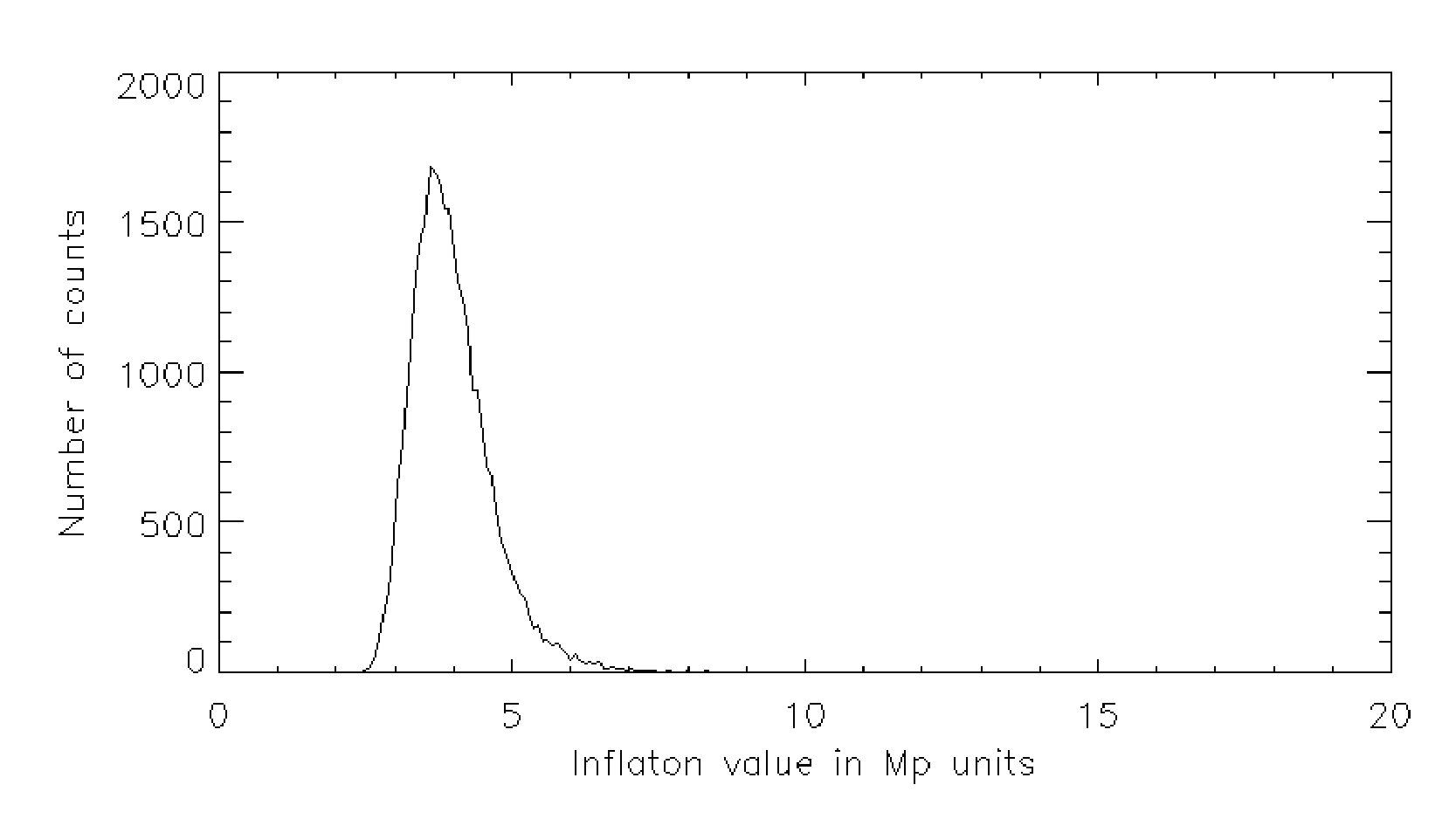}
\caption{Distribution of the estimated $\phi_*$ values for $r=0.001$.}
\label{fig:inflaton_dist1}
\includegraphics[width=8.6cm, height=7cm]{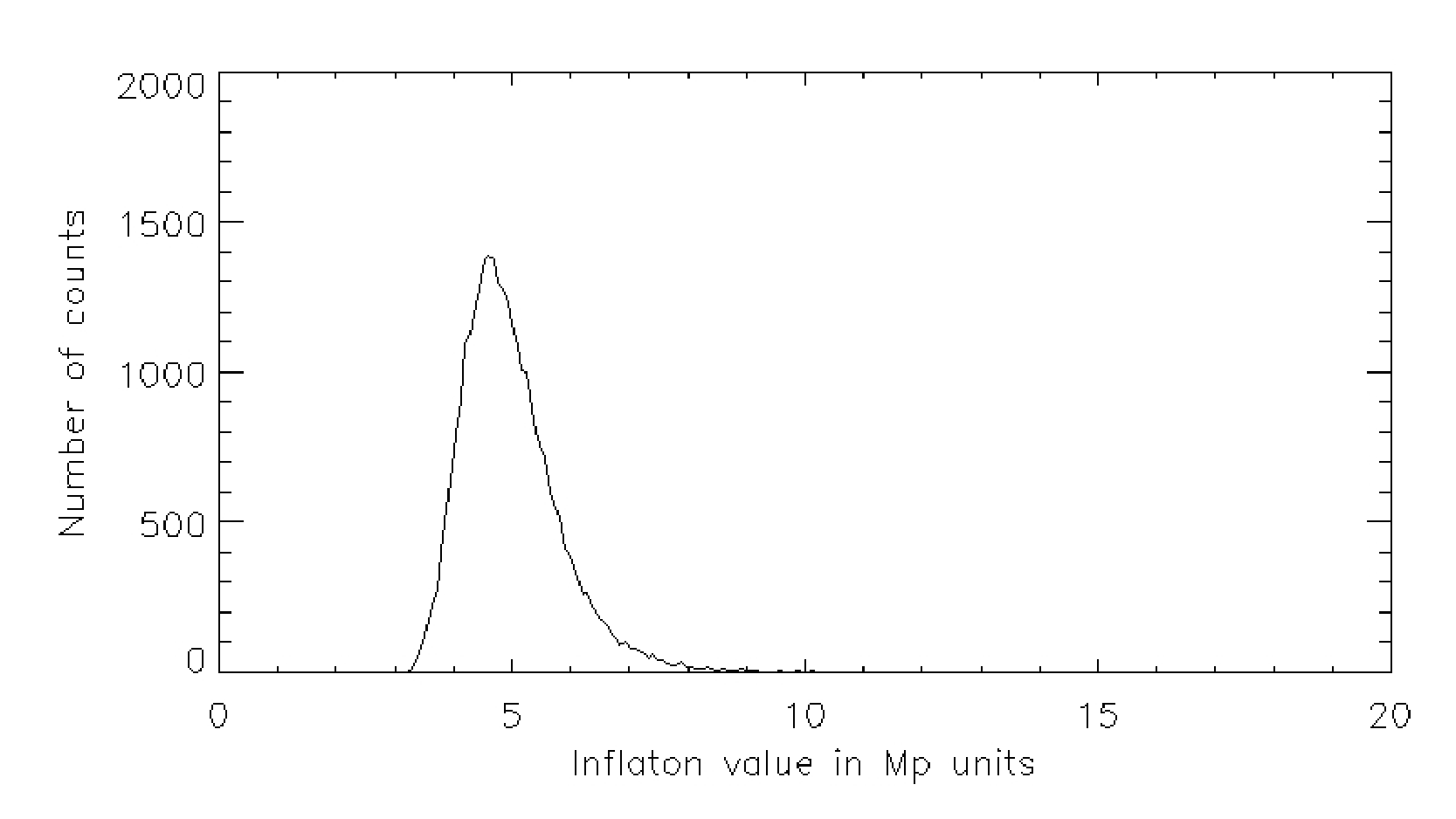}
\caption{Distribution of the estimated $\phi_*$ values for $r=0.002$.}
\label{fig:inflaton_dist2}
\includegraphics[width=8.6cm, height=7cm]{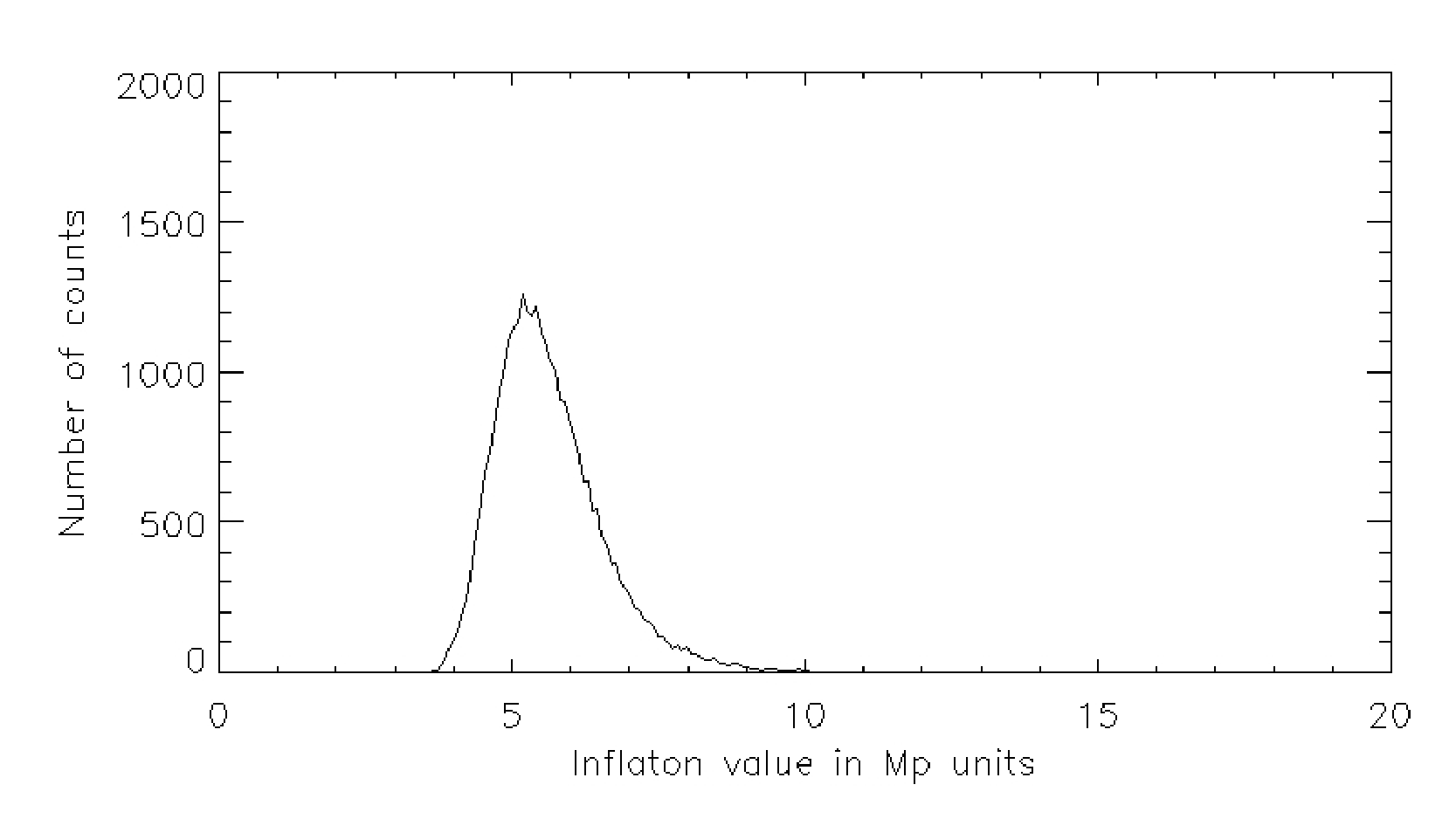}
\caption{Distribution of the estimated $\phi_*$ values for $r=0.003$. 
As one can expect, the vacuum expectation value of the scalar field increases as $r$ increase.}
\label{fig:inflaton_dist3}
\end{figure}
\begin{figure}[htbp]
\centering
\includegraphics[width=8.6cm, height=7cm]{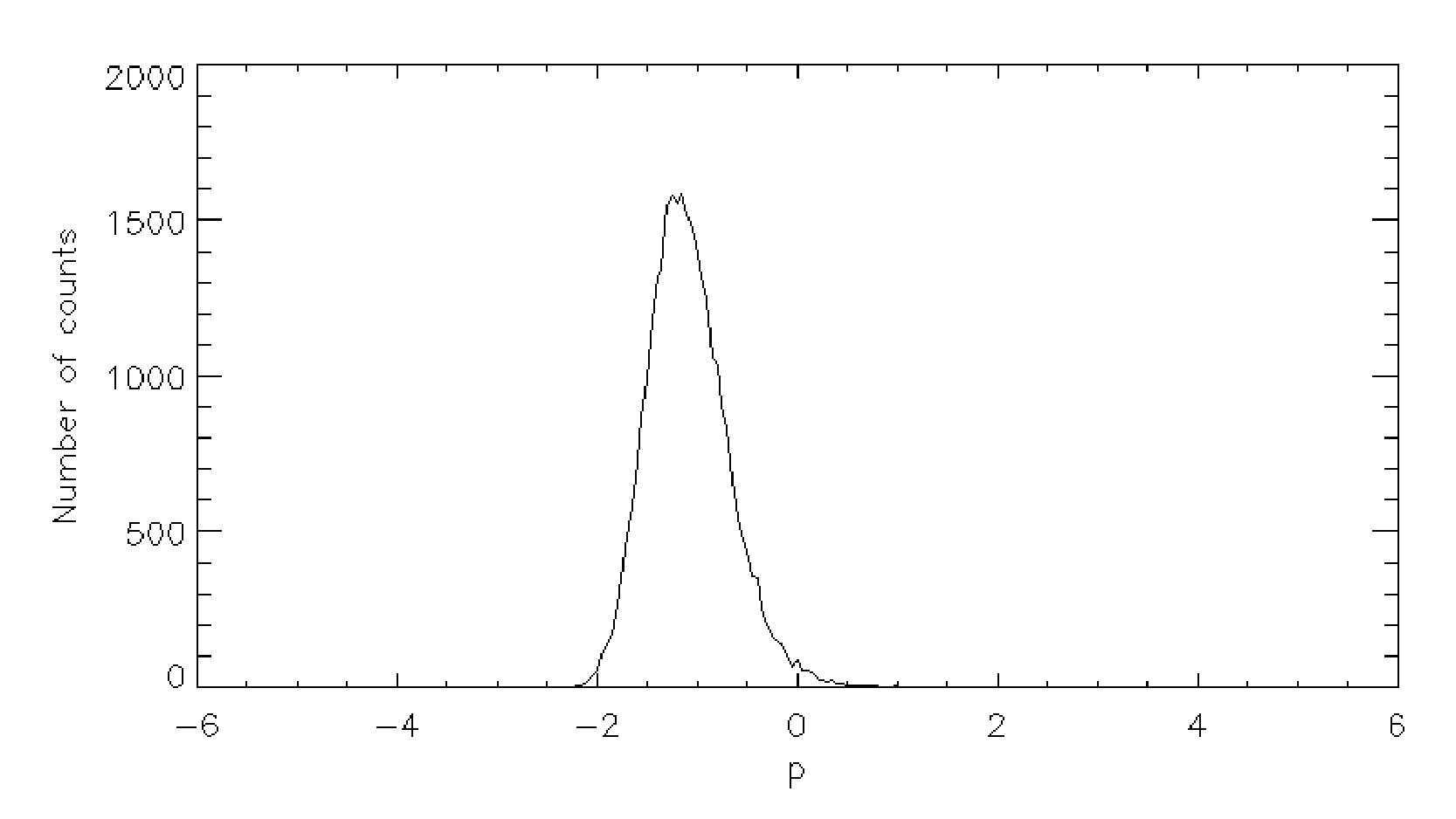}
\caption{Distribution of the $p=\ln\alpha$ values for $r=0.001$}
\label{fig:alfa_dist1}
\includegraphics[width=8.6cm, height=7cm]{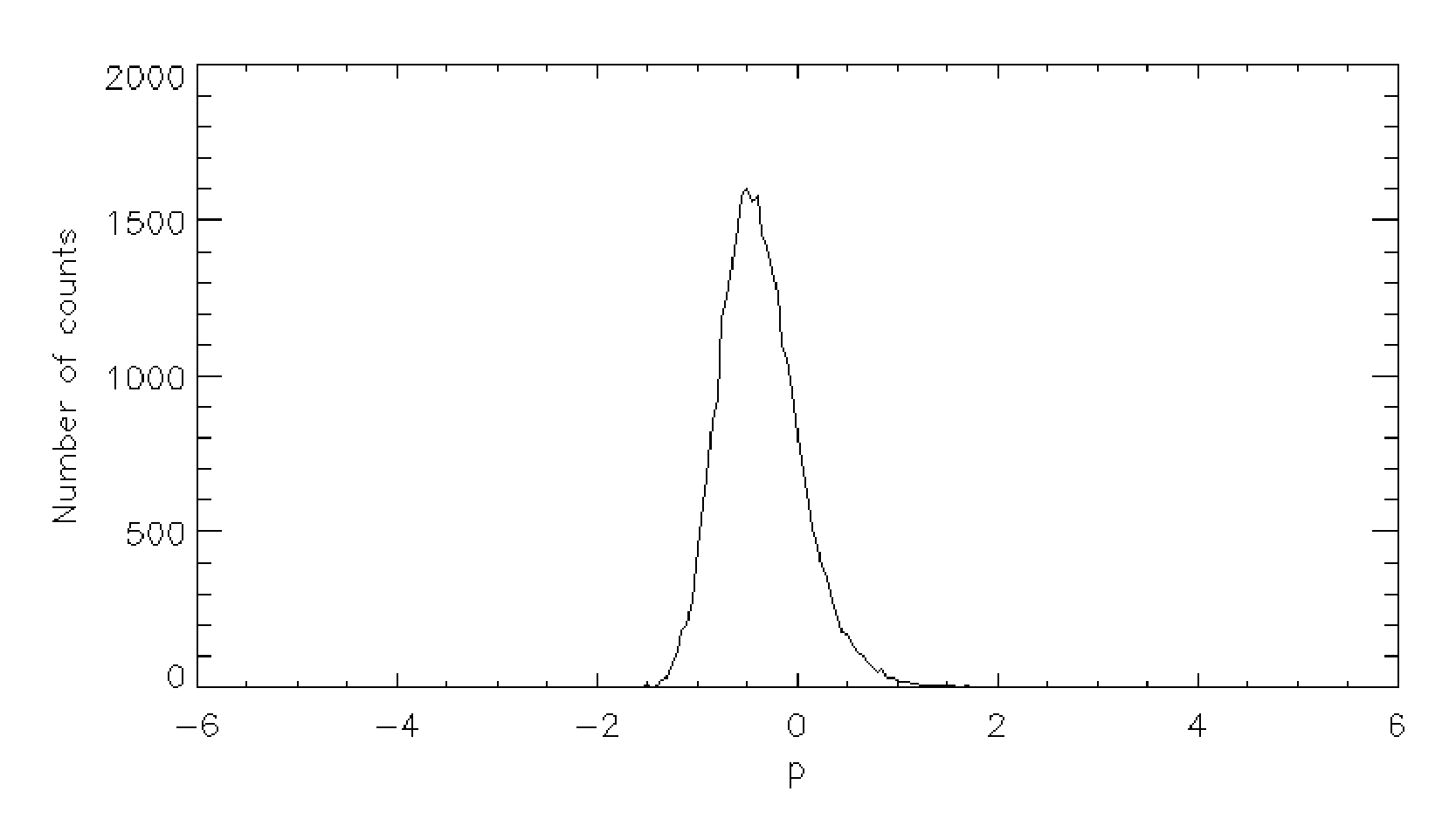}
\caption{Distribution of the $p=\ln\alpha$ values for $r=0.002$}
\label{fig:alfa_dist2}
\includegraphics[width=8.6cm, height=7cm]{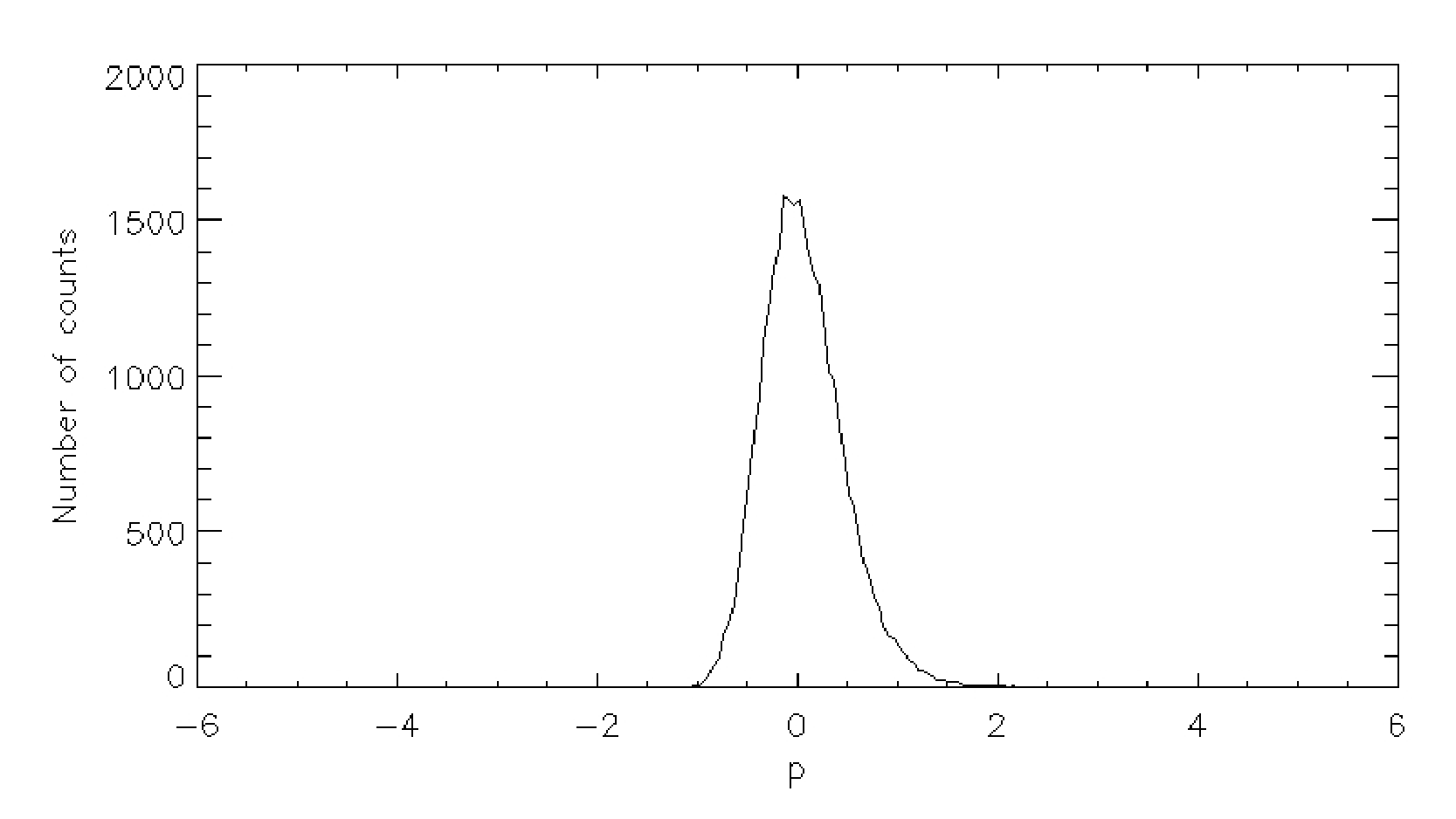}
\caption{Distribution of the $p=\ln\alpha$ values for $r=0.003$. 
The vacuum expectation value of $\alpha$ itself increases as $r$ increases, recovering Starobinsky inflation $\alpha\sim 1$ for $r=0.003$.}
\label{fig:alfa_dist3}
\end{figure}

The resulting distribution functions for $\phi_*$ are shown in Fig.$\ref{fig:inflaton_dist1}$, Fig.$\ref{fig:inflaton_dist2}$ and Fig.$\ref{fig:inflaton_dist3}$ while those of $p=\ln\alpha$ are shown in Fig.$\ref{fig:alfa_dist1}$, Fig.$\ref{fig:alfa_dist2}$ and Fig.$\ref{fig:alfa_dist3}$.
The mean and standard deviation values for the distribution of the coefficients $d_1$ and $d_2$ are summarized in Tab.I and in Tab.II.
The resulting mean and standard deviation values for the parameters $\phi_*$, $p$ and $R_K$ are summarized in Tab.III, Tab.IV and in Tab.V.
The Fig.$\ref{fig:inflaton_dist1}$, Fig.$\ref{fig:inflaton_dist2}$ and Fig.$\ref{fig:inflaton_dist3}$ and the Fig.$\ref{fig:alfa_dist1}$, Fig.$\ref{fig:alfa_dist2}$ and Fig.$\ref{fig:alfa_dist3}$ show the constraining power of a hypothetical CMB experiments of new generation.
As expected in this inflationary scenario, when the mean values of the tensor-to-scalar ratio increases, the mean values of $p$ and $\phi_*$ increase as well.
Indeed, the distribution functions of the constrained parameters move to high values in the frequency plots.
The mean value of $R_K$ decreases (in modulus) in complete agreement with its definition Eq.($\ref{eqn: kaler}$).

\section{Discussion and perspectives}

In the previous section we reconstructed, the probability distributions for the vev $\phi_*$, the parameter $\alpha$ and the scalar curvature $R_K$ for the supergravity $\alpha$-attractor models (E-model), starting from a set of observations for the main inflationary observables: the scalar spectral index, $n_s$,
and the tensor-to-scalar ratio, $r$.

\clearpage

\clearpage

\begin{figure}[htbp]
\centering
\includegraphics[width=8.6cm, height=6cm]{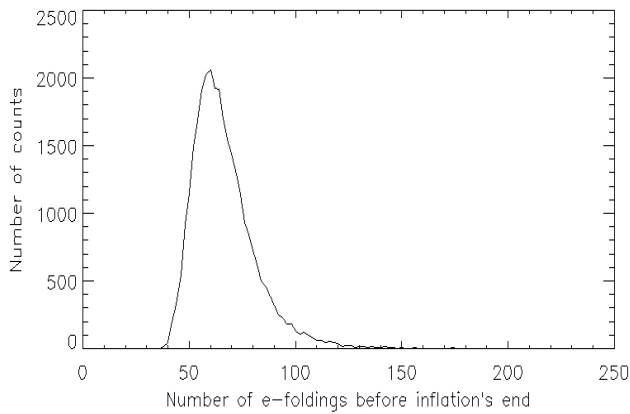}
\caption{Distribution function of the number of e-foldings before the end of inflation, $N_*$, related to the $\alpha$-attractor model with $r=0.003$ as a mean value of the tensor-to-scalar ratio. The result has been computed by the approximate solution of the classical equation of motion.}
\label{fig:efolds}
\end{figure}

The advantage of this method stands on the fact that just the computations of $V'$ and $V''$ are required.
Then, the ratio $c_1/c_2$ provides information on $\alpha$.
Note, as remarked in Sec. III, that in the usual blind approach to reconstruct the inflationary potential, the coefficients $d_1$ and $d_2$ given by
Eq.($\ref{eqn: coefficenti norm}$) seem do not depend on $\phi_*$. 
On the contrary, the main coefficients of the $\alpha$-attractor potential expansion $c_1$ and $c_2$ given by Eq.($\ref{eqn: c1}$),Eq.($\ref{eqn: c2}$)
are explicitly field dependent.
So, one could ask how our procedure (i.e, matching $c_1$ and $c_2$ with $d_1$ and $d_2$) is consistent with the general results of the blind expression. 
Actually, the consistency is guaranteed by the fact that the coefficients of the $\alpha$-attractor potential expansion $c_1$ and $c_2$ are evaluated at the moment of horizon exit as well as $d_1$ and $d_2$.
In fact, in the blind reconstruction, the expansion is again performed around $\phi_*$ but the dependence on $\phi_*$ is hidden in the choice of fixing ($n_s$,$r$) from  given experimental results and the $n_s$ and $r$ do depend on the value of the field (via slow roll parameters) as shown in Eq.($\ref{eqn: psrpobs}$) and Eq.($\ref{eqn: def}$) of Sec. II.
In our analysis it is clearly important to get an accurate estimate of $\phi_*$.
In fact, on the basis of Eq.($\ref{eqn: dep}$), one can infer that changing $N_*$ implies different values for $\phi_*$.
However, we can read this relation in the opposite sense: in each model of inflation, $N_*$ is sensitive to $\phi_*$ and to the other free parameters of the model, $\alpha$ in our case.
Therefore, one can conclude that an independent estimate of $\phi_*$ and $\alpha$
can provide information on $N_*$: $N_*=N_*(\phi_*,\alpha)$.
In this respect, it is also possible to put bounds on any deviations from the most common assumed value $N_*=60$.
Another important reason to explore this issue is that statistical information on $\alpha$ and $\phi_*$
may also imply information on postinflationary physics.
Indeed, distribution functions such as those previously reconstructed for $\alpha$ and $\phi_*$, can be used to provide a collection of possible post-inflationary energy density paths.
This can be done by solving numerically the appropriate system of coupled differential equations.
A discussion of this possibility is given in \cite{69}.
Moreover, $N_*$ (algebraically connected with $\phi_*$) depends on the physical events from inflation to recent epochs, i.e 
from a non trivial collection of quantities, 
as also summarized in \cite{69,70,71,72,73,74}. Usually, it is often expressed as
\begin{eqnarray}\label{eqn: efolds}
N_{*} &=& 67 - \ln \left(\frac{k_*}{a_0 H_0}\right) + \frac{1}{4} \ln\left( \frac{V^2_*}{M^4_p \rho_{end}}\right) \\ \nonumber
&&+ \frac{1-3w_{eff}}{12(1+w_{eff})} \ln\left(\frac{\rho_{reh}}{\rho_{end}}\right) - \frac{1}{12}\ln(g_{reh})  
\end{eqnarray}
Here, $a_0 H_0$ is the actual Hubble scale, $k_*$ is a pivot scale (typically of the order of $0.002$ Mpc$^{-1}$), $V_*=\Lambda^4$ is the inflationary energy scale, $\rho_{end}$ is the energy density at the end of inflation; $w_{eff}=p/\rho$ is
the effective equation of state of the reheating fluid, $\rho_{reh}$ is the energy density when reheating is completed and $g_{reh}$ is the effective number of boson degrees of freedom at $\rho_{reh}$.
However, from the complete expression of $N_*$, one can derive the number of e-foldings during reheating stage \cite{72,73,74}: 
\begin{eqnarray}\label{eqn: reh}
N_{reh}&=&\frac{4}{1-3 w_{eff}} \left[ -N_* - \ln\left( \frac{k_*}{a_0 H_0} \right) + \ln\left( \frac{T_0}{H_0} \right) \right] \\ \nonumber
&&+\frac{4}{1-3 w_{eff}}\left[ \frac{1}{4}\ln\left( \frac{V^2_*}{M^4_p \rho_{end}} \right) -\frac{1}{12}\ln(g_{reh})  \right] \\ \nonumber
&&+\frac{4}{1-3 w_{eff}}\left[ \frac{1}{4}\ln\left(\frac{1}{9}\right) + \ln\left( \frac{43}{11}\right)^{\frac{1}{3}} \left(\frac{\pi^2}{30} \right)^{\frac{1}{4}}\right]
\end{eqnarray}
So, once we know $N_*$ from Eq.($\ref{eqn: dep}$)
we can put better constraints on several aspects of the reheating physics by Eq.($\ref{eqn: reh}$). For example, we can derive distribution
function for the reheating temperature realized in the E-model $\alpha$-attractor framework, by the relation
\begin{eqnarray}
T_{reh}=\left( \frac{40 V_{end}}{\pi^2 g_{reh}} \right)^{1/4} \exp{ \left[-\frac{3}{4}(1+w_{eff})N_{reh}\right]}
\end{eqnarray}
An example of different reheating constraints can be found in \cite{72,73,74}.
In a forthcoming paper, we plan to apply the reconstruction technique presented above to describe the $\alpha$-attractor postaccelerated phase.\\
To conclude this last section, we want to discuss a ``naive" procedure to reconstruct the potential function. 
As we have seen in Secs. II and III, when a potential function is given, one can compute the related predictions in terms of the number of e-foldings.
In the case of $\alpha$-attractor models, at lowest order, one has
\begin{eqnarray}
n_s\sim1-\frac{2}{N_*}, \quad r\sim\frac{12 \alpha}{N_*^2}
\end{eqnarray}
However, starting from a given experimental or simulated CMB data set, we can reconstruct
the probability distributions for $\alpha$ and $N_*$, (instead of $\phi_*$).
This procedure is valid but the estimation of $N_*$ is related only to $n_s$ with no direct influence of $r$ that is (on the contrary) important to specify the attractor model and this influences the estimation of $\alpha$ or $b$:
\begin{eqnarray}\label{eqn: naive1}
N_*=\frac{2}{1-n_s}, \quad \alpha=\frac{r}{3 (1-n_s)^2}
\end{eqnarray}
In the standard reconstruction scheme, we do not pass through such a degree of approximation because of the Hamilton-Jacobi equation:
the extimation of $\alpha$ is more precise then the ``naive" case.
In fact, the limit of negligible $r$ in the Eq.($\ref{eqn: estb}$) corresponds to the $\alpha$-definition of Eq.($\ref{eqn: naive1}$).
Furthermore we also immediately provide an information on $\phi_*$.
Hereafter, we can constrain $N_*$ using the potential parameters and the vev of the field ($\alpha$, $\phi_*$) by the solution (approximeted or not)
of the equation of motion.
In Fig.$\ref{fig:efolds}$ we report the resulting distribution of $N_*$ in the case of $r=0.003$, using the first order solution of the classical equation of motion
given by Eq.($\ref{eqn: eqofmotion}$).

\section{Conclusions}

In this paper, we have used the potential reconstruction method to evaluate the inflaton field at horizon crossing and the potential parameter $\alpha$ of supergravity $\alpha$-attractor models. 
We have shown the possible constraints that next-generation CMB experiments can provide.
The method is applicable to different inflationary models and provides an estimate of $\phi_*$ completely independent by $N_*$.
In this sense, it is possible to use the vev as the input variable to estimate the number of e-foldings before the end of inflation and then, the postaccelerated physics.
There were two reasons for choosing E-models.
The first one lies in the capability of the model to interpolate a broad range of predictions for $r$.
The second one is related to the capability of these attractor models to reproduce during the accelerated phase, for some values of $\alpha$, the exponential potentials typically developed in the string inflation moduli context (see \cite{75,76,77,78,79} for detailed papers).		 

\begin{acknowledgments}

This work has been supported in part by the STaI ``Uncovering Excellence" grant of the University of Rome
``Tor Vergata", CUP E82L15000300005.
We are grateful to Nicola Bartolo, Sabino Matarrese and Dario Cannone for useful comments and discussions.
A.D.M also thanks Michele Cicoli for discussion on the recent developments of string theories.
P.C thanks L'isola che non c'\`e S.r.l for the support.

\end{acknowledgments}




\nocite{*}

\bibliography{apssamp}

\end{document}